\documentclass{aa}
\usepackage{graphicx}
\usepackage{longtable}
\usepackage{hyperref}
\usepackage{wasysym}
\usepackage{rotating}
\usepackage{natbib}
\usepackage{color}
\usepackage{lscape}
\usepackage{amsmath}
\usepackage{multirow}
\bibpunct{(}{)}{;}{a}{}{,} 
\newcommand{\Teff}   {$T_{\rm eff}~$}
\newcommand{\logg}   {$\log \mathrm{g}~$}
\newcommand{\feh}   {$[\mathrm{Fe/H}]~$}
\newcommand{\Teffs} {$T_{\rm eff}s~$}

\newcommand{\Teffe}   {$T_{\rm eff}$}
\newcommand{\logge}   {$\log \mathrm{g}$}
\newcommand{\fehe}   {$[\mathrm{Fe/H}]$}
\newcommand{\Teffse} {$T_{\rm eff}s$}

\begin{document}

\title{Fine structure of the age-chromospheric activity relation in solar-type stars\footnote{Table 1 is only available in electronic form
at the CDS via anonymous ftp to cdsarc.u-strasbg.fr (130.79.128.5)
or via http://cdsweb.u-strasbg.fr/cgi-bin/qcat?J/A+A/}\thanks{Based on spectroscopic observations
collected at the Observat\'orio do Pico dos Dias (OPD), operated
by the Laborat\'orio Nacional de Astrof\'{\i}sica, CNPq, Brazil,
and the European Southern Observatory (ESO), within the ON/ESO and
ON/IAG agreements, under FAPESP project n$^{\circ}$
1998/10138-8.}}

\subtitle{I. The Ca II infrared triplet: Absolute flux calibration}

\author{D. Lorenzo-Oliveira\inst{1,2}, G. F. Porto de Mello\inst{1}, L. Dutra-Ferreira\inst{1,3}, I. Ribas\inst{2}}

\offprints{D. Lorenzo-Oliveira, diego@astro.ufrj.br}

\institute{Observat\'orio
do Valongo, Universidade Federal do Rio de Janeiro, Ladeira do Pedro Antonio 43, CEP: 20080-090 Rio de Janeiro, RJ, Brazil\\
\email{diego@astro.ufrj.br,gustavo@astro.ufrj.br,leticia@astro.ufrj.br}
\and Institut de Ci\`encies de l'Espai (CSIC-IEEC), Facultat de Ci\`encies, Carrer de Can Magrans, s/n, Campus UAB, 08193 Bellaterra, Spain\\
\email{iribas@ice.cat} \and Departamento de Física Teórica e Experimental, Universidade Federal do Rio Grande do Norte, Campus Universitário Lagoa Nova, 59072-970, Natal, RN, Brazil \email{leticia@dfte.ufrn.br}}

\date{Received; accepted}

\authorrunning{Lorenzo-Oliveira et al.}

\titlerunning{Fine structure of the age-chromospheric activity relation in solar-type stars}

\abstract
{Strong spectral lines are useful indicators of stellar chromospheric activity. They are physically linked to the convection efficiency, differential rotation, and angular momentum evolution and  are a potential indicator of age. However, for ages > 2 Gyr, the age-activity relationship remains poorly constrained thus hampering its full application.} 
{The Ca II infrared triplet (IRT lines, $\lambda\lambda$ 8498, 8542, and 8662) has been poorly studied compared to classical chromospheric indicators. We report in this paper absolute chromospheric fluxes in the three Ca II IRT lines, based on a new calibration tied to up-to-date model atmospheres.}
{We obtain the Ca II IRT absolute fluxes for 113 FGK stars from high signal-to-noise ratio (S/N) and high-resolution spectra covering an extensive domain of chromospheric activity levels. We perform an absolute continuum flux calibration for the Ca II IRT lines anchored in atmospheric models calculated as an explicit function of effective temperatures (\Teffe), metallicity (\fehe), and gravities (\logge) avoiding the degeneracy usually present in photometric continuum calibrations based solely on color indices.}
%
{The internal uncertainties achieved for continuum absolute flux calculations are $\approx$ 2\% of the solar chromospheric flux, one order of magnitude lower than for photometric calibrations. Using Monte Carlo simulations, we gauge the impact of observational errors on the final chromospheric fluxes due to the absolute continuum flux calibration and find that \Teff uncertainties are properly mitigated by the photospheric correction leaving \feh as the dominating factor in the chromospheric flux uncertainty.}
{Across the FGK spectral types, the Ca II IRT lines are sensitive to chromospheric activity. The reduced internal uncertainties reported here enable us to build a new chromospheric absolute flux scale and explore the age-activity relation from the active regime down to very low activity levels and a wide range of \Teffe, mass, \fehe, and age.}


\maketitle

\section{Introduction}

The seminal work of \citet{skumanich72} established that stellar rotation in solar-type stars decays rapidly with age. The paradigm states that as an isolated star ages it loses a fraction of its mass through coronal winds. The mass loss leads consequently to a decrease in the angular momentum and the torque acts on the stellar surface slowing the rotation over millions of years. As a result, the stellar rotation braking causes a lower efficiency in the generation and amplification of magnetic fields at the base of the convective zone \citep{parker70} and less chromospheric heating \citep{noyes84}. As an observational indication of this effect, the profiles of strong spectral lines such as Ca II H \& K, infrared triplet lines, H$\alpha$, and other chromospheric indicators respond to changes in the temperature distribution at high atmospheric altitudes showing a reduced core chromospheric signature as the star ages. 

A number of empirical studies have attempted to convert the observed level of chromospheric emission into physical units \citep[][and references therein]{hall07}. This process enables a broad astrophysical interpretation of the chromospheric activity phenomena across a wide range of masses, ages, chemical compositions, and evolutionary states. The possibility of age-activity calibrations as well as their connections to rotational evolution, the properties of exoplanetary host stars, and chemodynamical evolution of the Galaxy turns the chromospheric activity estimates into important physical variables. In order to explore the nature of the chromospheric activity evolution, strong spectral features are commonly used. 

The lines of the Ca II infrared triplet at 8498.062 (T1), 8542.144 (T2), and 8662.170 (T3) \AA\, are formed in the lower chromosphere by subordinate transitions between the excited levels of Ca II $\mathrm{4^2P_{1/2,3/2}}$ and meta-stable $\mathrm{3^2D_{3/2,5/2}}$. This characteristic increases the opacity turning these spectral lines very intense and probing the physical conditions of a large range of atmospheric layers \citep{mihalas78}. Physically, it is expected that the chromospheric radiative losses of Ca II lines (Ca II IRT and H \& K) be strongly related because they share the same upper excited state ($\mathrm{4^2P_{1/2,3/2}}$) \citep{linsky79}. The consequence of the coupling is that these Ca II lines are mostly collisionally controlled in the lower chromosphere and very sensitive to the local temperature \citep{cauzzi08}. For these reasons, the Ca II IRT is studied as an indicator of chromospheric activity \citep{linsky79b,foing89,chmielewski00,busa07} revealing itself as an interesting alternative for solar-type chromospheres studies and angular momentum evolution \citep{krishnamurthi98}. These lines show attractive intrinsic characteristics such as:
\begin{enumerate}
   \item The stellar continuum in $\lambda\lambda$ 8400-8800 is barely affected by telluric lines \citep{busa07,dempsey93} and has a smaller density of photospheric lines which favors a more consistent spectra normalization and absolute continuum calibration of FGK stars in comparison to $\lambda\lambda$ 3900-4000. Furthermore, the theoretical flux distribution in the visible and infrared regions are better determined when compared to shorter wavelengths \citep{edvardsson08}.
    \item The profile of Ca II IRT lines (especially the wings) is sensitive to macroscopic fundamental parameters such as \Teffe, \fehe, and \logg \citep{andretta05}. These lines are important in the context of Galactic and extragalactic chemical enrichment as they allow the determination of Ca abundances or the estimation of [$\alpha$/Fe] in M dwarfs \citep{terrien15}.
\item Late-K and M stars have its flux distribution peaked towards longer wavelengths making the observation of visible and near-infrared chromospheric indicators such as H$\alpha$ and Ca II IRT easier when compared to UV counterparts like Ca II H \& K lines.
\item Their less pronounced contrast compared to other classical chromospheric indicators makes the Ca II IRT more suitable for the measurement of the stellar mean activity level since they are less sensitive to sudden modulations caused by flares and transient phenomena.
\end{enumerate}

A notorious disadvantage of the IRT lines is the lower instrumental contrast between the chromospheric and photospheric contributions compared to H$\alpha$ and H \& K lines, demanding spectra of good quality and a careful analysis of inactive stars. Moreover, \citep{andretta05} argue that the departures of LTE effects on the Ca II IRT become increasingly evident for lower \logg and lower \feh stars.  

There are different techniques for estimating chromospheric activity levels such as bolometric flux normalized indices, equivalent widths (EWs), and chromospheric absolute fluxes. The normalized indices \citep{noyes84,andretta05} and the use of equivalent widths \citep{busa07,zerjal13} have a non straightforward physical interpretation since these approaches are not ideal representations of the chromospheric radiative losses. 

The widely used $\log(R'_\mathrm{HK})$ Mount Wilson index, which is the ratio between chromospheric and bolometric fluxes \citep{noyes84}, relies on photometric \Teff estimates \citep{johnson66}. One of the disadvantages of this procedure is that color indices carry non-negligible additional dependencies of stellar chemical composition and evolutionary state. In order to take into account the latter effect, using PHOENIX models, \citet{mittag13} calibrated Ca II H \& K surface chromospheric flux levels for dwarfs, subgiants, and giants. 

Moreover, hidden intrinsic dependencies that manifest themselves in the spectral morphology are commonly disregarded in the literature. Together, these effects have important contributions to the interpretation of the age-activity relation and they must be well assessed quantitatively. Additionally, all Ca II IRT (near-infrared region) absolute flux calibrations in the literature are based on color indices \citep{linsky79b,hall96}. Such calibrations lack the necessary detail to bring out differences in the chromospheric fluxes caused by differences in stellar masses, metallicities, and evolutionary stages. 

These reasons motivate us to build a new absolute flux calibration for the near infrared continuum ($\approx$ $\lambda\lambda$ 8400-8750) in order to explicitly take into account the direct influences of \Teffe, \logge, and \fehe, using modern theoretical models of stellar atmospheres \citep{gustafsson08} which will enable us to perform a detailed study regarding the correlation between Ca II IRT chromospheric radiative losses and fundamental stellar parameters exploring a broad range of activity levels.

This paper is divided as follows. Sect. \ref{sec:obs_sample} describes the observations, stellar atmospheric, and evolutionary parameters derived, and the reduction of spectroscopic data. In Sect. \ref{sec:continuum_calib} we calculate the continuum absolute fluxes (erg cm$^{-2}$s$^{-1}$) as a function of atmospheric parameters (\Teffe, \logge, and \fehe) using LTE NMARCS models of atmospheres and we compare our results with photometric calibrations from the literature. In Sect. \ref{sec:total_abs_fluxes} we derive the total line absolute fluxes for T1, T2, and T3 lines. In Sect. \ref{sec:phot_correc} we perform the photospheric correction of the total absolute fluxes obtaining the estimates of Ca II IRT chromospheric radiative losses and discuss our results. The chromospheric flux errors analysis is discussed in Sect. \ref{sec:act_prob}. We advise those interested in the straightforward application of our method to follow the steps described in Sect. \ref{sec:steps}. Conclusions are drawn in Sect. \ref{sec:conclusions}.

\section{Sample stars and observations}\label{sec:obs_sample}

We observed 113 FGK dwarfs and subgiants in the near-infrared spectral region (NIR, $\approx$ 8300-8800 \AA) with high signal-to-noise ratio, covering a wide range of chromospheric activity levels. The NIR sample is composed of 95 main sequence (MS) and 23 subgiant (SG) stars limited to visual magnitude V = 11. Among them, potential members of kinematic groups (Ursae Majoris and HR1614) are present, as well as field stars and members of young open clusters such as the Pleiades and Hyades. In addition, we built a larger benchmark sample of 250 FGK stars containing all stars from the NIR sample. The observations of the benchmark sample were carried out around the H$\alpha$ region ($\approx$ 6480-6640 \AA) to determine precise spectroscopic \Teff for the entire NIR sample stars and, hence, establish a homogeneous absolute chromospheric flux scale.

Our NIR sample is composed of 84 stars observed with the FEROS high-resolution echelle spectrograph coupled to the 1.52 m telescope of ESO in La Silla (FEROS subsample) and 74 stars observed with the coud\'e spectrograph mounted at the 1.60m telescope of the Pico dos Dias Observatory (OPD, Braz\'opolis), hereafter called as OPD subsample. In order to obtain a homogeneous absolute flux scale, our total sample has 45 stars with both FEROS and OPD observations. 

The FEROS spectrograph has coverage from 3560 to 9200 \AA\ achieving a resolving power R = 48000. The spectra are automatically processed and calibrated in wavelength. The remaining reduction steps (Doppler correction and normalization) were performed in the conventional way. Due to the gap in echelle orders situated exactly around T2, it was not possible to estimate the chromospheric activity using this line for the FEROS data. The OPD sample covers all Ca II IRT lines achieving an intermediate resolving power of 18000. We adopted standard reduction steps (bias, flatfield, scattered light corrections, 1D extraction, Doppler correction and normalization). The range of S/N of our observations goes from 50 to 360 and the average is 150 $\pm$ 50 and 180 $\pm$ 60 for the FEROS and OPD samples, respectively. 

\begin{figure}[!htbp]
\centering
\begin{center}
  \begin{minipage}[h]{0.8\linewidth}
\resizebox{\hsize}{!}{\includegraphics{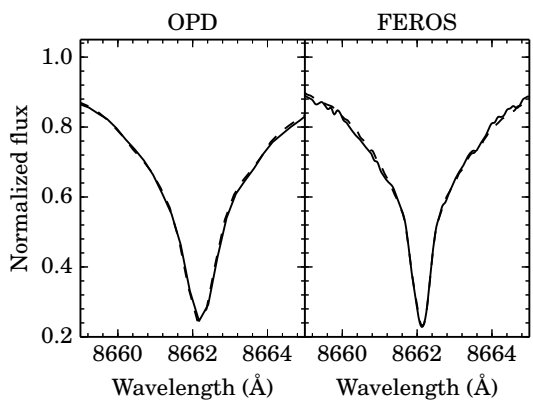}}
  \end{minipage}
\end{center}
\caption{OPD and FEROS subsamples with the two independently observed spectra of the same chromospherically inactive star over plotted. Left panel, the OPD spectra of HD 10700; right panel, the same for the FEROS spectra of HD 182572. The agreement between spectra observed in different runs is excellent for both spectrographs.}
\label{fig:normtest}
\end{figure}

We compared the spectra of stars with multiple observations in order to check consistency. In Fig. \ref{fig:normtest} the region around the center of the T3 Ca II triplet line is shown for two normalized spectra of the stars HD 182572 and HD 10700. Since these stars are quite old, absorption profiles are deep, showing low levels of activity, and no significant cycle modulation \citep{baliunas95,lovis11,hall07}. So, any difference was attributed to the reduction and normalization errors. It can be seen that both spectra show excellent agreement.

\subsection{H$\alpha$ sample}

Our data for the H$\alpha$ sample were obtained in several observational runs between 1994 and 2008. As the NIR sample, the spectra were collected using coudé spectrograph of the 1.60m OPD telescope. Over 14 years of observations, it was not possible to maintain the same instrumental setup, thus three different CCDs were used to collect our data: 2048$\times$4608, 2048$\times$2048 and 1024$\times$1024 pixels with 13$\mu$m, 13$\mu$m and 24$\mu$m of pixel size, respectively. We adopted a 1800 l/mm grating with 250$\mu$m slit-width centered at 6563\AA\, yielding a nominal resolution of 45000 for the 4608-pixel CCD and 20000 for the remaining ones\footnote{For a few stars with V > 10 from Pleiades open cluster and HR 1614 SKG, the slit-width was changed to 500$\mu$m yielding R = 10000}. In order to maintain internal consistency, we adopted the same reduction procedures as discussed above for the NIR sample. The typical S/N of this sample is 180 and the distribution ranges from 40 to 420 wherein 90\% of them have S/N > 100 \citep[part of this sample was published in][]{lyra05}.     

We obtained spectroscopic \Teff determinations for all stars of the H$\alpha$ sample by fitting theoretical line profiles to the observed ones, following exactly the same prescriptions of Lyra \& Porto de Mello (2005). Full details are given by these authors and here we merely summarize the essential aspects. The damping wings of the H$\alpha$ line are classical \Teff indicators \citep{gehren81,barklem02,lyra05} for solar-type stars, being strongly \Teff sensitive but showing little sensitivity to surface gravity, metallicity, microturbulent velocity, and NLTE effects. On the other hand, due to the breadth of the line wings, continuum placement is a significant source of error in the \Teff determination. The use of echelle spectra thus hampers proper use of the method since a precise continuum normalization around H$\alpha$ demands a consistent blaze correction over adjacent orders \citep{ramirez14,barklem02}, which is at best very difficult and often impossible. We therefore took advantage of our OPD single order spectra to obtain a precise and homogeneous normalization around the H$\alpha$ profile and hence internally very consistent \Teffse. The stellar metallicities were taken from assorted sources in the literature. To further enhance the internal consistency of the stellar atmospheric parameters we applied an empirical zero-point correction to the literature metallicites, adopting the H$\alpha$ Teff scale as homogeneous and correcting the \feh values by $\Delta$\feh/$\Delta$\Teff = -0.06 dex/100 K, where $\Delta$\Teff is the difference between the literature \Teff and the H$\alpha$ \Teff obtained here.

As an additional consistent \Teff determination, photometric \Teffs were derived for all stars in the H$\alpha$ benchmark sample from the calibrations of \citet{portodemello14}, taking into account explicitly the corrected \feh values. The $\mathrm{(B-V)}$, $\mathrm{(B_T-V_T)}$, and $\mathrm{(b-y)}$ color indices were used, taken from the Hipparcos and the Olsen catalogues \citep{olsen83,olsen93,olsen94}: all (b-y) photometry was converted to the \citet{olsen93} scale according to prescriptions given by this author. The straight average between the H$\alpha$ and the photometric \Teff determinations was adopted as the \Teff scale for the present work.

\begin{figure}[!htbp]
\begin{center}
  \begin{minipage}[h]{0.9\linewidth}
\resizebox{\hsize}{!}{\includegraphics{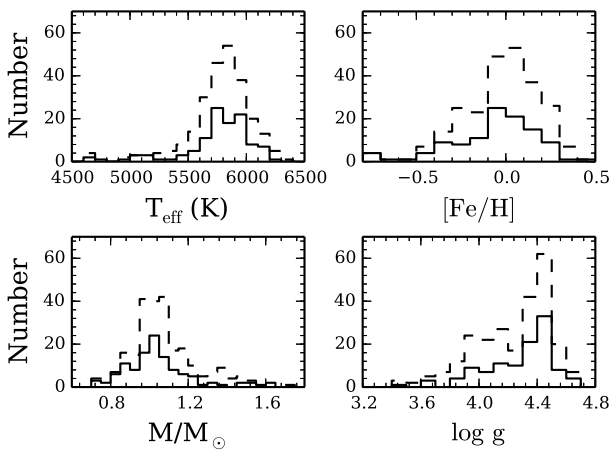}}
  \end{minipage}
\end{center}
\caption{Distributions of \Teffe, \fehe, mass, and \logg of H$\alpha$ and NIR samples represented by dashed and solid lines, respectively.}
\label{fig:histparam}
\end{figure}

From Hipparcos parallaxes \citep{vanleeuwen07}, and the \Teffe, and \feh values derived above, we calculated stellar luminosities using bolometric corrections from \citet{flower96}. Masses and surface gravities were obtained from theoretical evolutionary tracks due to \citet{kim02} and \citet{yi03}. We adopt this homogeneous scale of evolutionary surface gravities as our logg scale. Since it has recently become clear that the structural and temporal evolution of chromospheric activity is modulated by other parameters besides stellar age \citep{rochapinto98,lyra05,mamajek08}, we purposefully buit a sample of field dwarfs and subgiants covering an extensive range of metallicities (from $\approx$ -0.8 to +0.4 dex), and masses (from $\approx$ 0.7 to 1.5 M$_{\odot}$). The atmospheric parameters \Teffe, \feh derived by us are given in Table \ref{table:atmparam} for the full H$\alpha$ benchmark sample, also containing all of the stars considered for the construction of the Ca II IRT lines chromospheric indicator. The internal typical uncertainties for \Teff (mean of H$\alpha$ and photometric \Teffse), \feh (from the literature but corrected for the homogeneous H$\alpha$ \Teff scale), and \logg (evolutionary) are, respectively, 50 K, 0.07 dex, and 0.1 dex. In Fig. \ref{fig:histparam} we show the distribution of stellar parameters derived in this work.
\begin{table*}[!htbp]
\begin{scriptsize}
\begin{center}
\begin{tabular}{c | c c c c | c c c c | c}
\hline
\hline
HD & \Teff(K) & \feh & \feh & References & \Teff(K) & \Teff(K) & \Teff(K) & $\sigma$ (K)& NIR Sample \\
- & Literature & Literature & Corrected & - & H$\alpha$ & Photometric & Adopted & - & - \\
\hline
1461 & 5724 & 0.17 & 0.22 & \citet{ramirez12} & 5803.0 & 5788.6 & 5795.8 & 50 & Yes \\
1581 & 5922 & -0.21 & -0.21 & \citet{ramirez12} & 5929.0 & 6000.4 & 5964.7 & 50 & Yes \\
1835 & 5817 & 0.21 & 0.23 & \citet{ramirez12} & 5846.0 & 5837.5 & 5841.8 & 50 & No \\
2151 & 5816 & -0.12 & -0.09 & \citet{ramirez12} & 5863.0 & 5935.3 & 5899.1 & 50 & Yes \\
3795 & 5385 & -0.61 & -0.54 & \citet{ramirez13} & 5506.0 & 5431.0 & 5468.5 & 50 & Yes \\
3823 & 5981 & -0.35 & -0.46 & \citet{ramirez12} & 5802.0 & 5968.0 & 5885.0 & 50 & No \\
4307 & 5785 & -0.25 & -0.29 & \citet{ramirez12} & 5723.0 & 5854.6 & 5788.8 & 50 & No \\
4308 & 5720 & -0.29 & -0.30 & \citet{portodemello14} & 5695.0 & 5713.5 & 5704.3 & 50 & No \\
4391 & 5878 & -0.03 & -0.06 & \citet{santos04} & 5829.0 & 5839.4 & 5834.2 & 50 & Yes \\
7570 & 6087 & 0.15 & 0.17 & \citet{ramirez12} & 6122.0 & 6087.6 & 6104.8 & 50 & Yes \\
...  & ...  & ...  & ...  & ...  & ...  & ...  & ...  & ...  & ...  \\
  \hline
\end{tabular}
\end{center}
\caption{Atmospheric parameters collected from the literature and derived in this work. The full table is available online.}
\label{table:atmparam}
\end{scriptsize}
\end{table*}

\section{Absolute flux calibration} \label{sec:continuum_calib}
\subsection{Total line flux equation}
An important procedure in our analysis is the calculation of the emergent flux from the observed star whose relative scale is normalized by a pseudo-continuum in all spectra. Therefore, we do not have direct observational access to absolute purely radiative losses arising from stellar chromospheres. We then relate the observed $f$ and intrinsic flux $\mathcal {F}$ of a given star and integration bandwidth ($\Delta \lambda$) by the well-known relation: 
\begin{equation}\label{f6}
\mathcal{F}_{L} (\Delta \lambda_{L}) = \frac{f_{L} (\Delta \lambda_{L})}{f_{C} (\Delta \lambda_{C})} \mathcal{F}_{C} (\Delta \lambda_{C}),
\end{equation}
where:
\begin{equation}\label{f7}
\mathcal{F}_{C}(\Delta\lambda_{C}) = \overline{F} \Delta \lambda_{C}.
\end{equation}

Eq. \ref{f6} is extremely important for our purposes because it relates directly to the absolute flux at the center of a specific spectral line $\mathcal{F}_{L}$ through an empirical scaling factor given by the ratio of observed fluxes ($\frac{f_ {L} (\Delta\lambda_{L})} {f_{C} (\Delta\lambda_{C})}$) which is multiplied by the absolute flux at the stellar continuum ($\mathcal{F}_{C}(\Delta\lambda_{C})$) in each reference region. Thus, we avoid the theoretical difficulties that arise from modeling the complex NLTE effects which are present at the center of intense spectral lines such as the Ca II IRT lines \citep{andretta05} leaving only the absolute flux in the stellar continuum to be theoretically estimated by LTE models. The scale factor is observationally chosen in the observed spectra and is a function of the spectral resolution. Numerical integration is straightforwardly performed for the line and the continuum bands. 

\subsection{NMARCS atmospheric models}
We calculated the theoretical continuum absolute fluxes using 1D, hydrostatic, plane-parallel LTE NMARCS spectra \citep{gustafsson08} with a constant resolving power (R = $\lambda/{\Delta\lambda}$ = 20000, 900 \AA\ to 20000 \AA) along all the range covered by our observations, a value consistent with the resolution of our observed spectra (8340 \AA\ to 8750 \AA, R = $\lambda/{\Delta\lambda} \approx$ 18000 for the OPD data). 

\begin{figure}[!htbp]
\begin{center}
  \begin{minipage}[h]{0.8\linewidth}
\resizebox{\hsize}{!}{\includegraphics{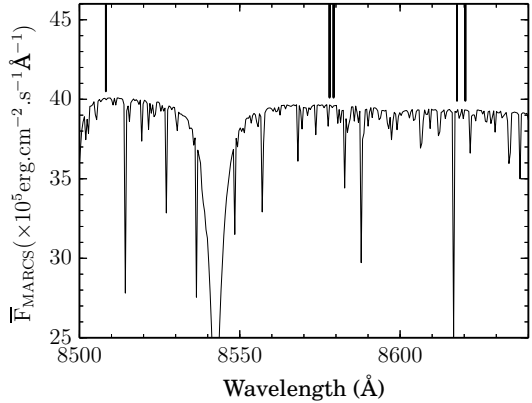}}
  \end{minipage}
\end{center}
\caption{Theoretical flux distribution model for a star cooler than the Sun (\Teff = 5500 K, \feh = 0.0 e \logg = 4.4 dex). The vertical lines indicate the central wavelength of the adopted reference regions. The large transition seen is the T2 line at $\lambda$8542.}
\label{fig:NMARCS2}
\end{figure}

Based on the solar spectrum catalog of Utrecht \citep{moore66} and also the Solar Flux Atlas of \citet{kurucz84}, we identified a total of 5 continuum reference regions listed in Table \ref{table:regselecionadas}. The identification of each RR is considerably straightforward since the density of photospheric transitions is quite low compared to shorter wavelengths. As an example, we show in Fig. \ref{fig:NMARCS2} the theoretical spectrum of a solar-type star (\Teff = 5500 K, \logg = 4.4, and \feh = 0.0 dex), identifying the central wavelength of the reference regions by solid vertical lines. 
\begin{table}[!htbp]
\begin{center}
\begin{tabular}{c c c c c}
  \hline
 Adopted Reference Regions (\rm{RR})  & Wavelength (\AA) \\
\hline
\rm{RR}$_1$ & 8507.50 to 8509.00 \\
\rm{RR}$_2$ & 8577.50 to 8581.00\\
\rm{RR}$_3$ & 8577.50 to 8578.50 \\
\rm{RR}$_4$ & 8617.00 to 8618.50 \\
\rm{RR}$_5$ & 8619.70 to 8621.00 \\
  \hline
\end{tabular}
\end{center}
\caption{Selected RR for the total absolute flux calculation.}
\label{table:regselecionadas}
\end{table}
\begin{figure}[!htbp]
\begin{center}
  \begin{minipage}[h]{1\linewidth}
\resizebox{\hsize}{!}{\includegraphics{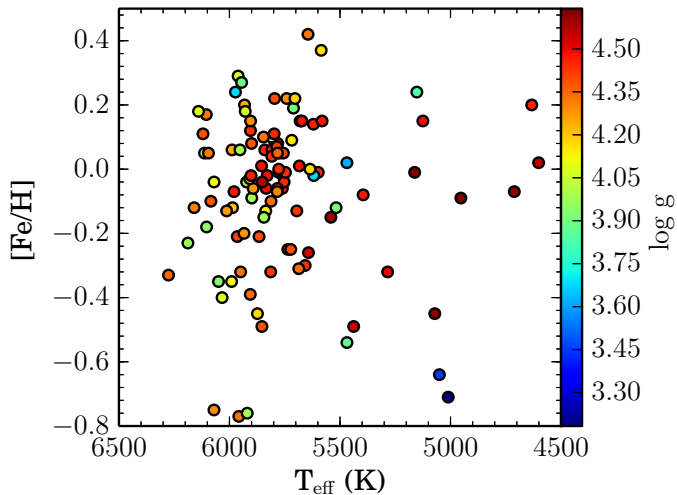}}
  \end{minipage}
\end{center}
\caption{Distribution of \Teff, \feh for the sample stars, with the \logg color coded. The atmospheric parameters of the entire sample is bracketed by the our NMARCS model atmosphere grid.}
\label{fig:NMARCS3}
\end{figure}

The theoretical atmospheric model parameters range in \Teff from  4500 K to 6500 K in steps of 250 K; the surface gravities (cgs units) from 3.0 to 5.0 dex in steps of 0.5 dex; and \feh from -1.0 to 0.5 dex in steps of 0.25 dex. For each reference region, we interpolate the fluxes $\overline{F}$ for specific values of \Teffe, \logge, and \fehe. The models completely cover the parameter space of our sample. 

\subsection{Absolute continuum flux calibration: $\overline{F}$}\label{ss:integration}
We next derive a calibration enabling the calculation of the model stellar absolute continuum flux ($\overline{F}$) for each reference region as a function of the observational atmospheric parameters \Teffe, \logge, and \fehe. Various consistency tests with regressive models indicated that a cubic model shows neither departures from a normal distribution nor correlation between residuals and the model fluxes. The functional form adopted was the third-order polynomial given by:  
\begin{equation}\label{mcub}
\begin{split}
\overline{F} (x,y,z) = \beta_0 + \beta_1x + \beta_2y + \beta_3z + \beta_4xy + \beta_5xz \\ + \beta_6yz  + \beta_7x^2  + \beta_8y^2 + \beta_9z^2 + \beta_{10} x^2y  \\ + \beta_{11} x^2z  + \beta_{12} y^2 x + \beta_{13} y^2 z + \beta_{14} z^2 x \\ + \beta_{15} z^2 y  + \beta_{16}xyz +\beta_{17}x^3 \\ + \beta_{18}y^3 + \beta_{19}z^3,
\end{split}
\end{equation}
where $x$ = \Teffe/5777, $y$ =  \logge, $z$ = \fehe, and $\overline{F} (x,y,z)$ is in $\times$ 10$^5$ erg cm$^{-2}$ s$^{-1}$\AA$^{-1}$ units.

\begin{figure}[!hbp]
\begin{center}
  \begin{minipage}[t]{0.49 \linewidth}
    \resizebox{\hsize}{!}{\includegraphics{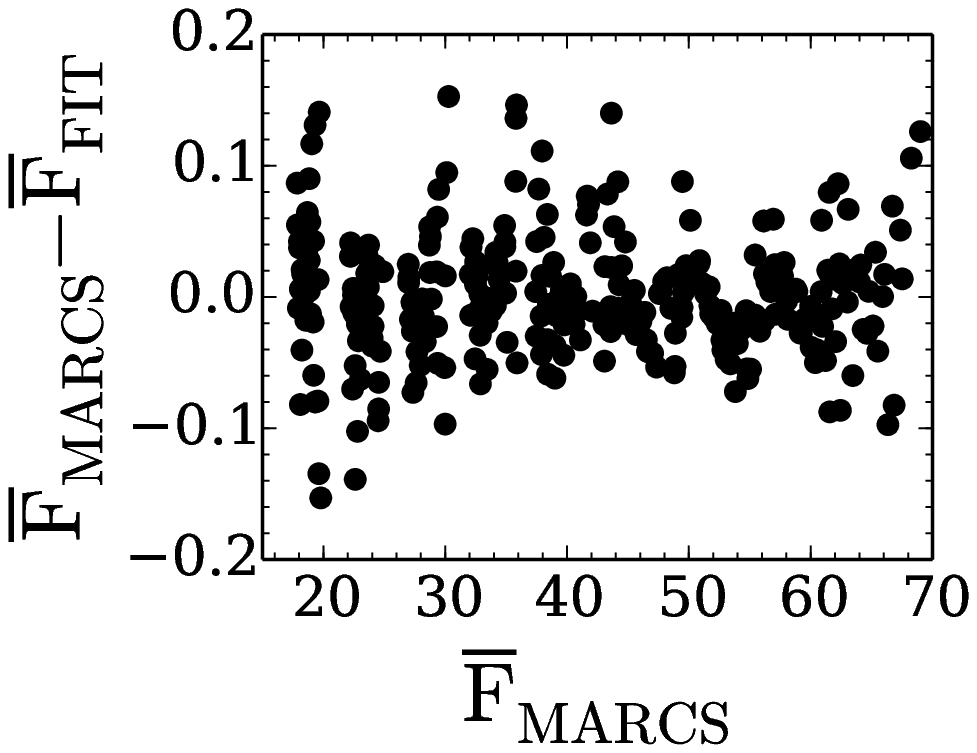}}
  \end{minipage}
  \begin{minipage}[t]{0.49 \linewidth}
    \resizebox{\hsize}{!}{\includegraphics{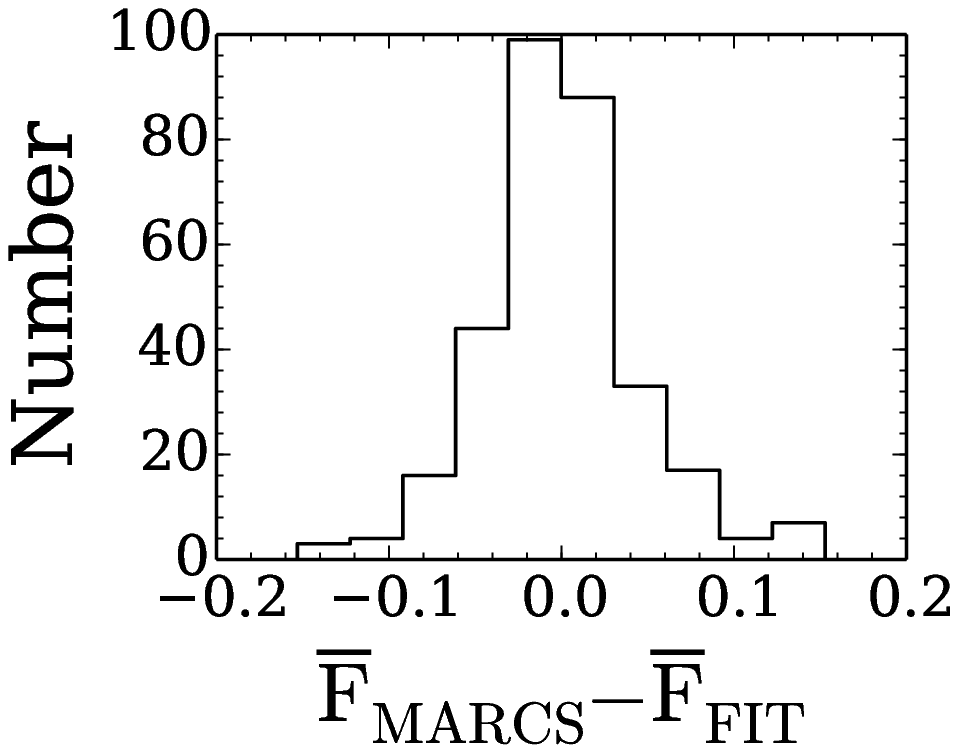}}
  \end{minipage}
\end{center}
\caption{Distribution (left panel) and histogram (right panel) of RR1 residuals for the cubic fitting model. \textit{Left panel}: The differences between the NMARCS theoretical fluxes and the fluxes obtained from our regression models as a function of the NMARCS theoretical fluxes. \textit{Right panel}: The distribution of the residuals from the relation showed in the left panel. The absolute continuum flux is in 10$^5$ erg cm$^{-2}$ s$^{-1}$\AA$^{-1}$ units.}
\label{RM_cubico}
\end{figure}

We adopted an advanced variable selection procedure (\textit{Stepwise regression}). A detailed discussion on this procedure is given by \citet{ghezzi14}, and here we provide only the essencial aspects. For each reference region, we initialy fit a third-order polynomial and test iteratively the statistical significance of each term according to its hierarchy, starting from the highest order term. We then remove only the terms which decrease the previous fitting error ($\sigma_{\mathrm{FIT}}$) and the BIC index \citep[Bayesian Information Criteria,][]{kass95}. The iteractions continue until there are no possible removals leading to a final reduced model with the same predictive power of the complete ones. We provide, in Table \ref{table:coeff}, all the relevant information about the regressive model of absolute flux of each reference region. We see that in all RR that had removed variables, the $\Delta$BIC lies between 4.7 and 15.9 indicating that the reduced models are more suitable than the complete ones \citep{kass95}. In Fig. \ref{RM_cubico}, we show the distribution of the residuals for RR1. The typical standard deviation found is 0.05 $\times$ 10$^5$ erg cm$^{-2}$ s$^{-1}$\AA$^{-1}$. 

\begin{table*}[!htbp]
\begin{footnotesize}
\caption{Coefficients of the reduced regressive models. First and second columns are the coefficients of the model describing the reference regions and additional information regarding the first column. The remaining columns show, for each RR, the values of the corresponding coefficient described in first and second columns (following Eq. \ref{mcub}). Respectively, the last four rows are: fitting errors in 10$^5$ erg cm$^{-2}$ s$^{-1}$\AA$^{-1}$ units, R$^2$ multiple correlation coefficient, BIC difference between complete, and final reduced cubic regressive models, and the variables that were removed during the \textit{Stepwise} regression. The percentages that are shown together with the fitting errors are its fraction compared to typical solar chromospheric radiative losses derived in Sect. \ref{sec:phot_correc}. $RR_3$ did not suffer any removal, reason why the BIC difference is zero.}
\label{table:coeff}
\begin{center}
\begin{tabular}{c c c c c c c}
  \hline
$\beta$ & Additional Information & RR1 & RR2  & RR3 & RR4 & RR5   \\
  \hline
$\beta_0$ & intercept & 39.211 & 46.545 & 45.791 & 38.112 & 43.060\\
$\beta_1$ &$\mathrm{T_{eff}}$ & -214.985 & -215.697 & -219.870 & -200.457 & -197.401\\
$\beta_2$ & $\log\mathrm{ g }$& 4.350 & -0.195 & 0.518 & 0.591 & -2.013\\
$\beta_3$ &$\mathrm{[Fe/H]}$ & -0.444 & 0.964 & -3.285 & -6.774 & -1.320\\
$\beta_4$ &$\mathrm{T_{eff}}\log\mathrm{ g }$ & -12.713 & 7.492 & 6.000 & 3.889 & 8.912\\
$\beta_5$ &$\mathrm{T_{eff}}\mathrm{[Fe/H]}$ & 5.651 & -2.778 & 5.654 & 9.476 & 3.672\\
$\beta_6$ &$\log\mathrm{ g }\mathrm{[Fe/H]}$ & -2.035 & -1.044 & -1.049 & -0.715 & -1.287\\
$\beta_7$ &$\mathrm{T_{eff}}^2$ & 325.152 & 274.256 & 285.581 & 272.625 & 253.623\\
$\beta_8$ &$\log^2\mathrm{ g }$ & -0.003 & -0.972 & -0.979 & -0.705 & -0.692\\
$\beta_9$ & $\mathrm{[Fe/H]}^2$ & -0.832 & -1.151 & -1.688 & -2.791 & -1.022\\
$\beta_{10}$ &$\mathrm{T_{eff}}^2\log\mathrm{ g }$ & 7.427 & -8.709 & -8.170 & -6.970 & -9.380\\
$\beta_{11}$ &$\mathrm{T_{eff}}^2\mathrm{[Fe/H]}$ & -3.282 & 4.068 & 0.963 & 0.000 & 1.359\\
$\beta_{12}$ &$\mathrm{T_{eff}}\log^2\mathrm{ g }$ & 0.170 & 1.050 & 1.082 & 1.025 & 1.005\\
$\beta_{13}$ &$\log\mathrm{ g }^2\mathrm{[Fe/H]}$ & -0.057 & -0.040 & -0.033 & -0.025 & -0.030\\
$\beta_{14}$ &$\mathrm{T_{eff}}\mathrm{[Fe/H]}^2$ & 1.805 & 1.804 & 3.009 & 3.339 & 2.656\\
$\beta_{15}$ &$\log\mathrm{ g }\mathrm{[Fe/H]}^2$& 0.000 & 0.000 & -0.062 & 0.000 & -0.102\\
$\beta_{16}$ &$\mathrm{T_{eff}}\log\mathrm{ g }\mathrm{[Fe/H]}$ & 2.698 & 1.470 & 1.297 & 0.976 & 1.455\\
$\beta_{17}$ &$\mathrm{T_{eff}}^3$ & -101.789 & -56.074 & -61.606 & -59.915 & -48.618\\
$\beta_{18}$ &$\log^3\mathrm{ g }$ & 0.000 & 0.019 & 0.018 & 0.000 & 0.000\\
$\beta_{19}$ &$\mathrm{[Fe/H]}^3$ & 0.116 & 0.000 & 0.143 & -0.117 & 0.220\\
\hline
$\sigma_{\mathrm{FIT}}$& $\times$ 10$^5$ erg cm$^{-2}$ s$^{-1}$\AA$^{-1}$& 0.048(1.2\%) & 0.059(1.2\%) & 0.065(1.5\%) & 0.073(1.8\%) & 0.068(1.7\%)\\
R$^2$ & & 1.0 & 1.0 & 1.0 & 1.0 & 1.0\\
$\Delta BIC$ & BIC(M) - BIC(MR) & 11.2 & 11.1 & 0.0 & 15.9 & 4.7\\
Removed Coefficients & &  $\beta_{18}$,$\beta_{15}$ & $\beta_{19}$,$\beta_{15}$ & none & $\beta_{18}$,$\beta_{15}$,$\beta_{11}$ & $\beta_{18}$ \\
   \hline
\end{tabular}
\end{center}
\end{footnotesize}
\end{table*}

We found \Teff to be the major responsible for the variance of theoretical absolute continuum flux ($\approx$ 90\%), as expected, therefore transfering to all terms that include it a large statistical significance. The other variables (\logg and \fehe) combined account for $\approx$ 10\%. For instance, we calculated the total continuum absolute fluxes for each RR, obtaining $\approx$ 4 $\times 10^6$ erg cm$^{-2}$ s$^{-1}$ which represents 3 orders of magnitude larger than $\sigma_{\mathrm{FIT}}$ (see Table \ref{table:coeff}). We expect the chromospheric component of a very inactive Sun-like star to be $\approx$ $10^{5}$ erg cm$^{-2}$ s$^{-1}$ (see, for example, Figs. \ref{fig:term} and \ref{fig:dist_act}), 1-2 orders of magnitude higher than our fitting uncertainties. So, the absolute continuum calibration should not represent any significant additional source of error in the age-activity relations. 

\begin{figure}[!hbp]
\begin{center}
  \begin{minipage}[h]{0.7 \linewidth}
\resizebox{\hsize}{!}{\includegraphics{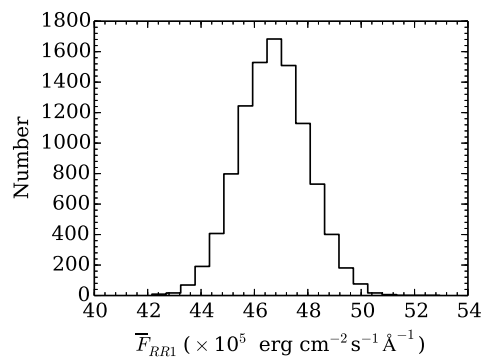}}
  \end{minipage}
\end{center}
\caption{Impact of observational errors in the predicted absolute flux values of theoretical atmospheric models for the solar atmospheric parameters. We generate a distribution of absolute continuum flux based on 10$^4$ MC simulations assuming gaussian uncertainties of 50 K, 0.1 dex, and 0.07 dex for \Teffe, \logge, and \fehe, respectivelly. The typical standard deviation found in this distribution is $\approx$ 1.3 $\times$ 10$^5$ erg cm$^{-2}$ s$^{-1}$.}
\label{erros_obs}
\end{figure}

In Fig. \ref{erros_obs}, we show the impact of observational errors in the absolute continuum flux distribution, $\overline{F}_C$ (\Teffe, \logge, \fehe). We generated $10^4$ Monte Carlo simulations for $\overline{F}_C$ assuming a Gaussian distribution of the observational uncertainties of atmospheric parameters ($\sigma_{\mathrm{T_{eff}}}$ = 50 K, $\sigma_{\log\mathrm{ g }}$ = 0.1 dex, $\sigma_{\mathrm{[Fe/H]}}$ = 0.07 dex). We calculated the standard deviation ($\sigma_{\mathrm{parameters}}$) of the total flux distribution shown in Fig. \ref{erros_obs} for RR1 and found typical values of 1.3 $\times$ 10$^5$ erg cm$^{-2}$ s$^{-1}$\AA$^{-1}$ ($\approx$ 30\% of solar chromospheric fluxes). When we propagate only \Teff errors in absolute fluxes, the flux variance turns out to be 1.2 $\times$ 10$^5$ erg cm$^{-2}$ s$^{-1}$\AA$^{-1}$, which corresponds to approximately 95\% of $\sigma_{\mathrm{parameters}}$, showing the great importance of the accurate determination of this parameter. However it must be emphasized that these errors are strongly correlated with \Teffe, which means that higher \Teff leads to higher total line fluxes independent of the intrinsic chromospheric activity level. We expect to mitigate this residual correlation isolating the chromospheric component by a proper photospheric correction procedure. We will discuss the details of this procedure in Sect. \ref{sec:phot_correc} and \ref{sec:act_prob}.

\subsection{Comparison with the literature: \rm{[Fe/H] bias}} \label{ss:metbias}

\citet{hall96}, hereafter H96, calibrated absolute continuum flux estimates ($\overline{F}_{H96}$) for luminosity classes I-V covering the near ultraviolet (Ca II H \& K) up to near infrared (Ca II IR triplet) region as a function of different color indices. To compare this with our results, we focus in his NIR calibration which is the region around T1 and T2 lines ($\lambda$8520). Depending on the color indices adopted, the errors derived from his functional relation were $\approx$ 10\% which corresponds to $\approx$ 5 $\times$ 10$^5$ erg cm$^{-2}$ s$^{-1}$\AA$^{-1}$ in the solar case (\Teff = 5777 K and \feh = 0.0). At this magnitude, the fitting errors will be of the same order of the chromospheric radiative losses of a typical inactive sun-like star.

\begin{figure}[!htbp]
\begin{center}
  \begin{minipage}[t]{0.8 \linewidth}
    \resizebox{\hsize}{!}{\includegraphics{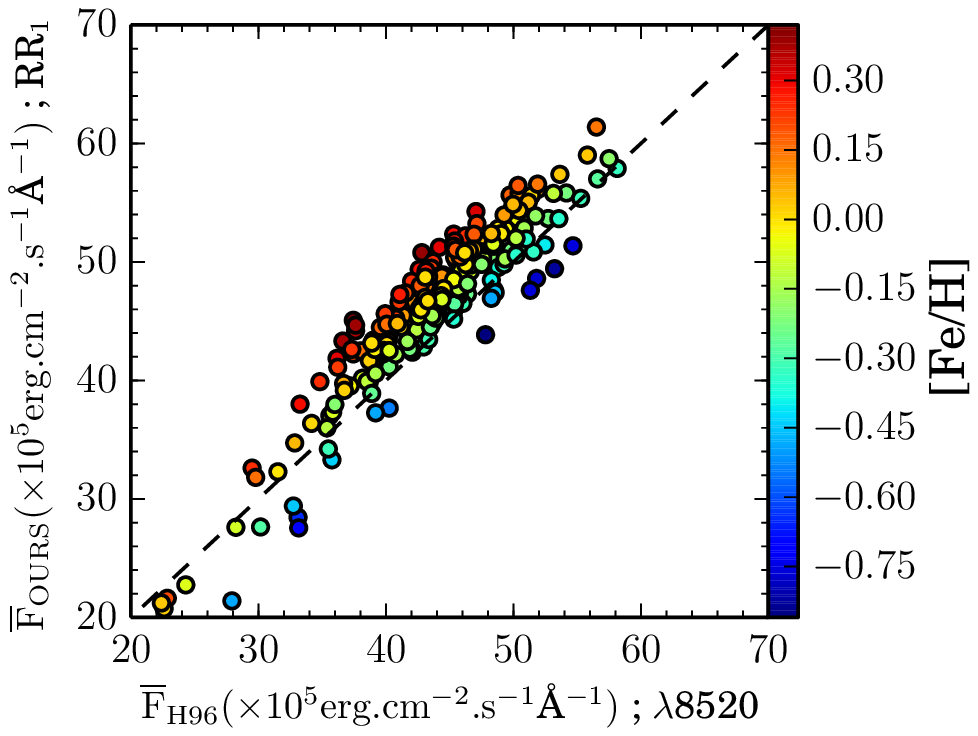}}
  \end{minipage}
  \begin{minipage}[t]{0.8 \linewidth}
    \resizebox{\hsize}{!}{\includegraphics{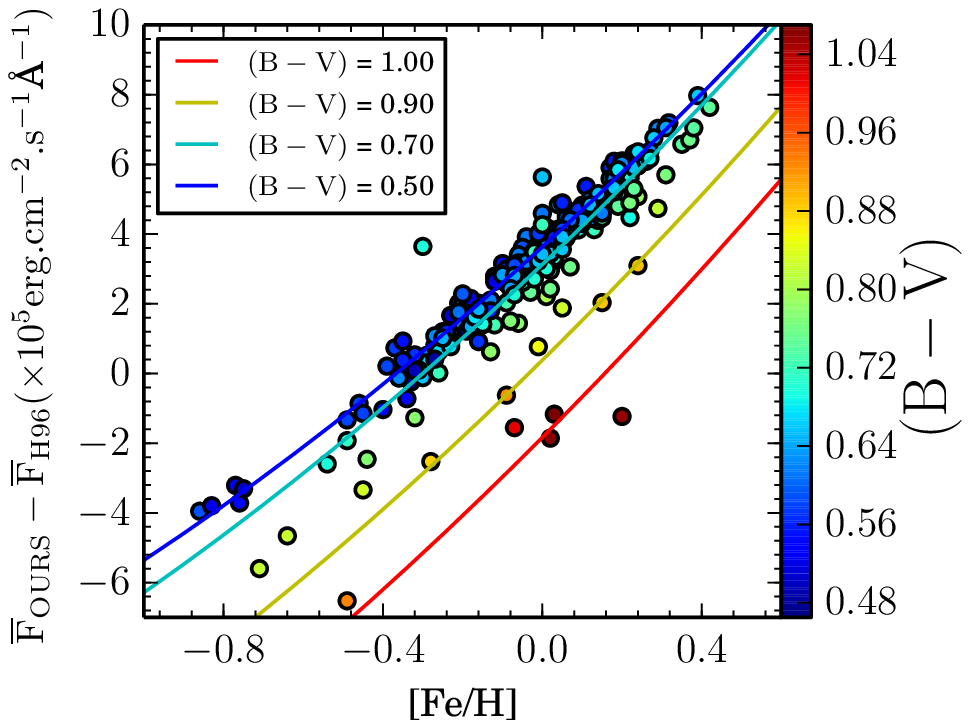}}
  \end{minipage}
\end{center}
\caption{\textit{Upper panel:} Comparison between \citet{hall96} and our calibration of near-IR continuum absolute fluxes. \textit{Lower panel:} Strong \feh correlation in the differences between the absolute continuum flux estimates. The colors are related to different \feh(upper panel) and $\mathrm{(B-V)}$ (lower panel) estimates for each star. The solid lines are correction functions for specific $\mathrm{(B-V)}$ indices.}
\label{fig:hall_flux}
\end{figure}

In Fig. \ref{fig:hall_flux}, we show the comparison between our RR$_1$ fluxes calculated for the 250 stars with H$\alpha$ \Teff in our benchmark sample and the $\lambda$8520 fluxes based on the (B-V) color index calibration taken from H96. Visually, the excellent agreement between the absolute continuum flux scales could be confirmed by a very strong linear correlation $\rho$ = +0.93 (\textit{Spearman's} correlation). The \feh effects appear to be the most important that distinguish one relation from the other since a clear and systematic linear pattern could be detected as we analyzed the differences between the two determinations as a function of \feh (right panel). For a given \Teffe, metal-rich stars have more efficient blocking of continuum radiation and, consequently, the temperature gradient of its deeper layers will be enhanced leading to an extra heating and flux \citep[e.g.~][]{edvardsson93}.

We stress two points: 1) In the lower panel, the standard deviation of the differences between continuum fluxes is $\approx$ 3 $\times$ 10$^5$ erg cm$^{-2}$ s$^{-1}$\AA$^{-1}$ which is close to the fitting errors derived by H96 considering color indices; 2) Considering the typical levels of chromospheric radiative losses for stars after the Vaughan Preston Gap ($\geq$ 2 Gyr), the \feh effects must have growing importance as we consider stars progressively more inactive. This result reinforces the need for a new absolute continuum calibration that has flexibility to distinguish the \fehe, \Teffe, and \logg continuum effects. Probably, for stars with $\mathrm{(B-V)}$ $\leq$ 0.75, just a simple correction depending on \feh is enough to bring the \cite{hall96} near-IR continuum fluxes to values closer to ours. However, for cooler stars, the \feh effect seems to be correlated with $\mathrm{(B-V)}$ as well, demanding a slightly more complex correction $\kappa$(\feh,$\mathrm{(B-V)}$):
\begin{equation}
\overline{F}_{RR1} = \overline{F}_{H96} + \kappa(\mathrm{[Fe/H]},\mathrm{(B-V)}),
\end{equation}
where:
\begin{equation}
\begin{split}
\kappa(\mathrm{[Fe/H]},\mathrm{(B-V)}) = - 4.94 + 9.36\mathrm{[Fe/H]} + 31.16\mathrm{(B-V)} \\ + 2.13\mathrm{[Fe/H]}\mathrm{(B-V)} + 1.44\mathrm{[Fe/H]}^2 \\ - 28.06\mathrm{(B-V)}^2
\end{split}
\end{equation}
and the internal uncertainty of the flux conversion is $\sigma$ = 0.29 $\times$ 10$^5$ erg cm$^{-2}$ s$^{-1}$\AA$^{-1}$.

\section{Absolute total line fluxes: $\mathcal{F}_{L} (\Delta \lambda_{L})$} \label{sec:total_abs_fluxes}

Returning to Eq. \ref{f6}, we calculated the term $\frac{f_{L} (\Delta\lambda_{L})} {f_{C} (\Delta\lambda_{C})}$ by numerical integrations leaving the integration bandwidth $\Delta\lambda_{L}$ around the Ca II IRT line cores a free parameter to be determined.

\begin{figure}[!htbp]
\begin{center}
  \begin{minipage}[t]{1\linewidth}
    \resizebox{\hsize}{!}{\includegraphics{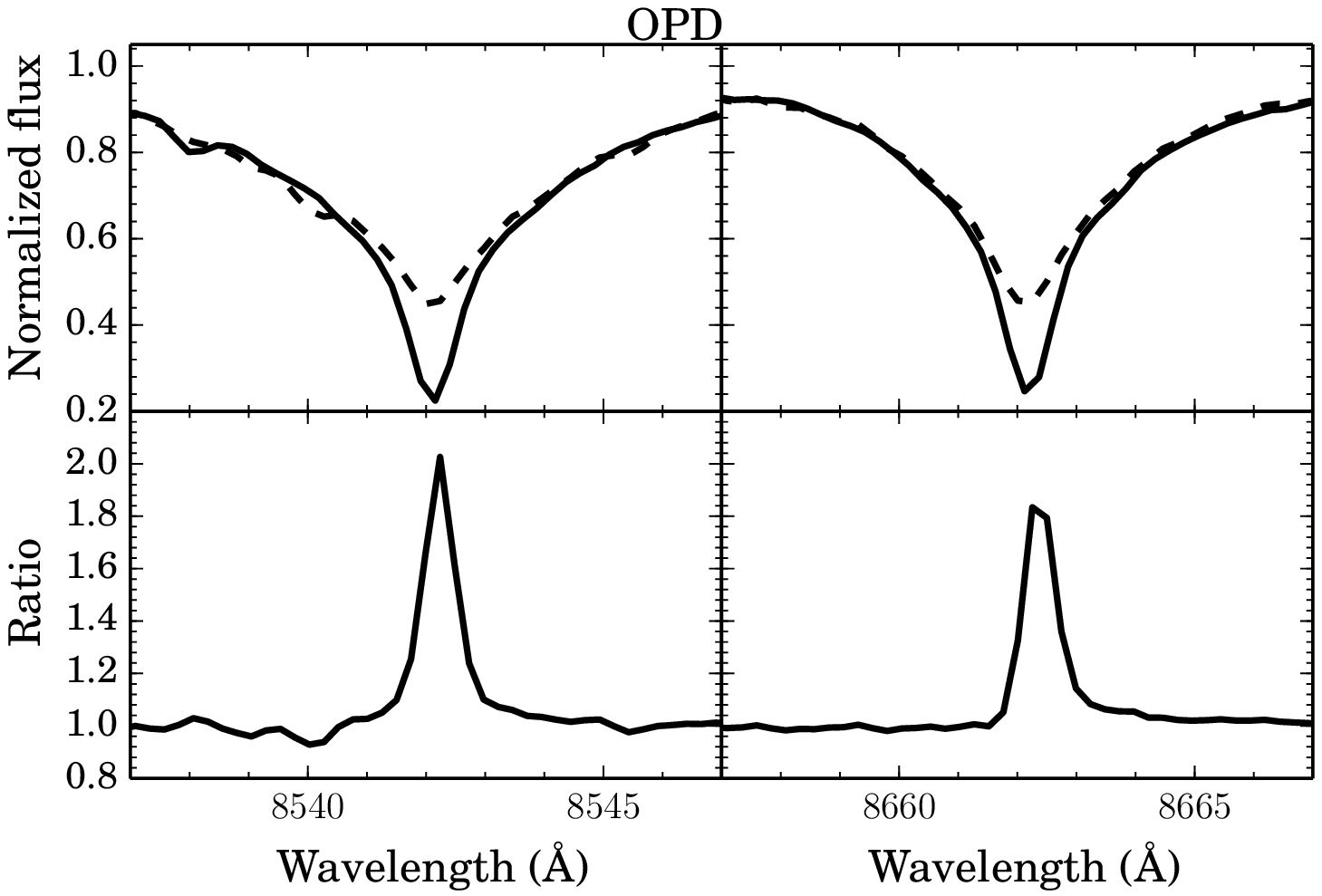}}
  \end{minipage}
  \begin{minipage}[t]{1\linewidth}
    \resizebox{\hsize}{!}{\includegraphics{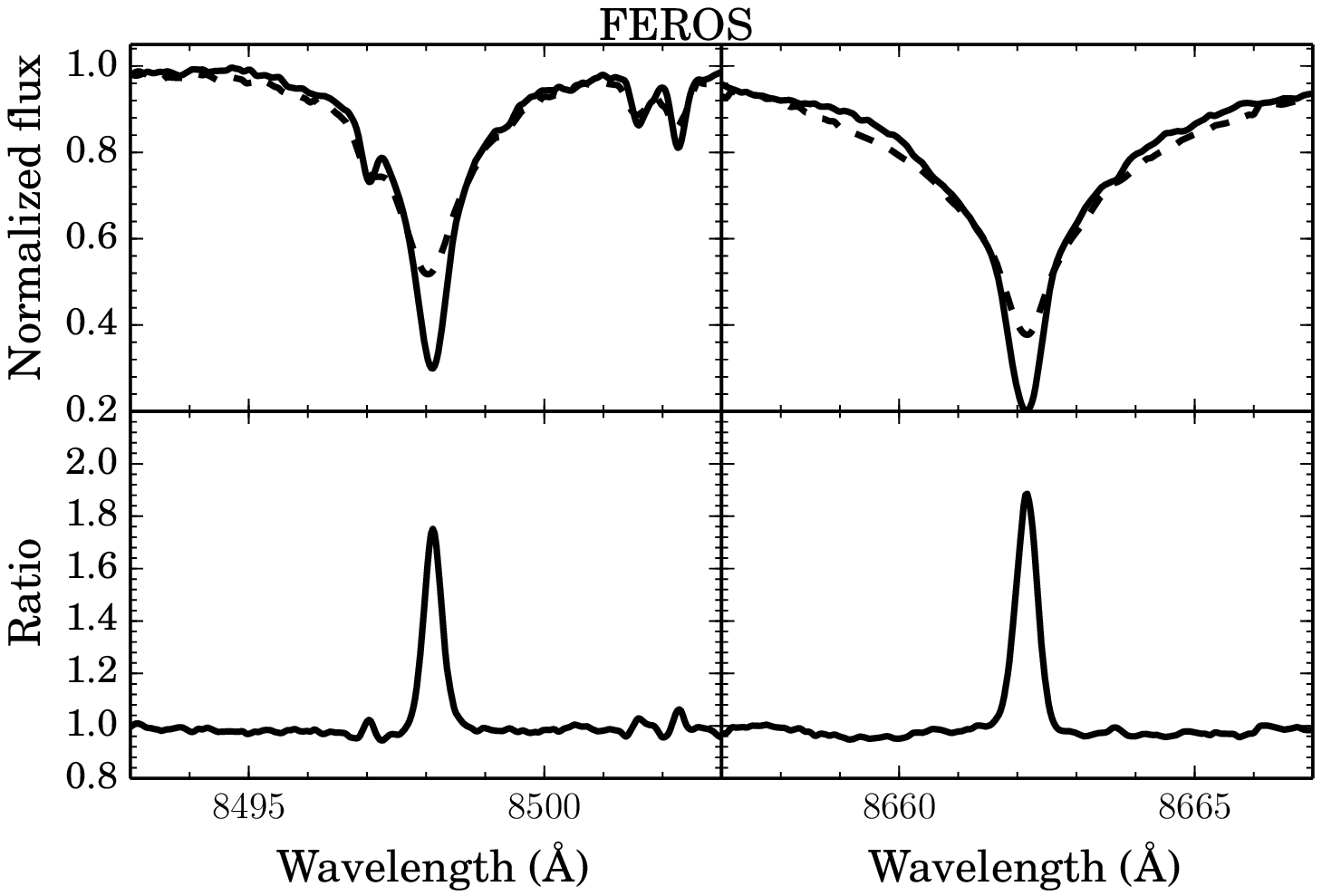}}
  \end{minipage}
\end{center}
\caption{Two selected ratio spectra from the FEROS and OPD samples. We choose pairs composed of an active (HD 28992 for the OPD and HD 165185 for FEROS samples) and an inactive star (HD 2151 for both samples). The profile differences in the line cores are mostly due to the chomospheric activity component.}
\label{razao1}
\end{figure}

In Fig. \ref{razao1} we show the normalized and ratio spectra of chromospherically active and inactive stars. These were chosen to possess similar atmospheric parameters in order to isolate the chromospheric activity differences. We identify two distinct regimes connected by a smooth transition. The first and more obvious one shows a sharp flux ratio contrast, suggesting that at lower optical depths (higher altitudes) there is an additional physical mechanism (chromospheric) to the second regime (photospheric) represented by an approximately constant flux ratio. 

Given the problems linked to the arbitrariness on determining the exact interval around the center of Ca II IRT lines where the chromospheric component is predominant without too much influence of thermal processes, we chose pairs of similar stars composed of an active and inactive star and calculated their total absolute fluxes. In the absence of any a priori suitable initial value of the width of the interval of integration, we covered all possible values of $\Delta\lambda_L$ over an extensive domain ranging from bandwidths which evidently have a high contrast in total  absolute flux rate (0.8 \AA) to the upper limit (4.0 \AA) where the width is clearly dominated by the photospheric component. In successive numerical integration, we adopt steps of 0.05 \AA\ in order to ensure a continuous visual behavior, which facilitates interpretation regarding the variation of the total absolute fluxes ratio ($\mathcal{R}$) as function of the integration bandwidth ($\Delta\lambda_L$).
From these calculations, for each pair of stars, we calculated the first order gradient ($\nabla\mathcal{R}$) and second order ($\nabla^2 \mathcal{R}$) of this ratio represented mathematically by the set of equations
\begin{equation}
\langle \mathcal{F}_{L} (\Delta \lambda_{L}^j) \rangle = \frac{1}{5}\sum_{i=1}^{i=5}\frac{f_{L} (\Delta \lambda_{L}^j)}{f_{C} (\Delta \lambda_{C}^i)}\mathcal{F}_{C}(\Delta\lambda_{C}^i),
\end{equation}
\begin{equation}
\mathcal{R}_j = \frac{\langle \mathcal{F}_{L} (\Delta \lambda_{L}^j) \rangle^{Active}}{\langle \mathcal{F}_{L} (\Delta \lambda_{L}^j) \rangle^{Inactive}}
\end{equation}

\begin{figure}[!htbp]
  \begin{minipage}[h]{0.49\linewidth}
\resizebox{\hsize}{!}{\includegraphics{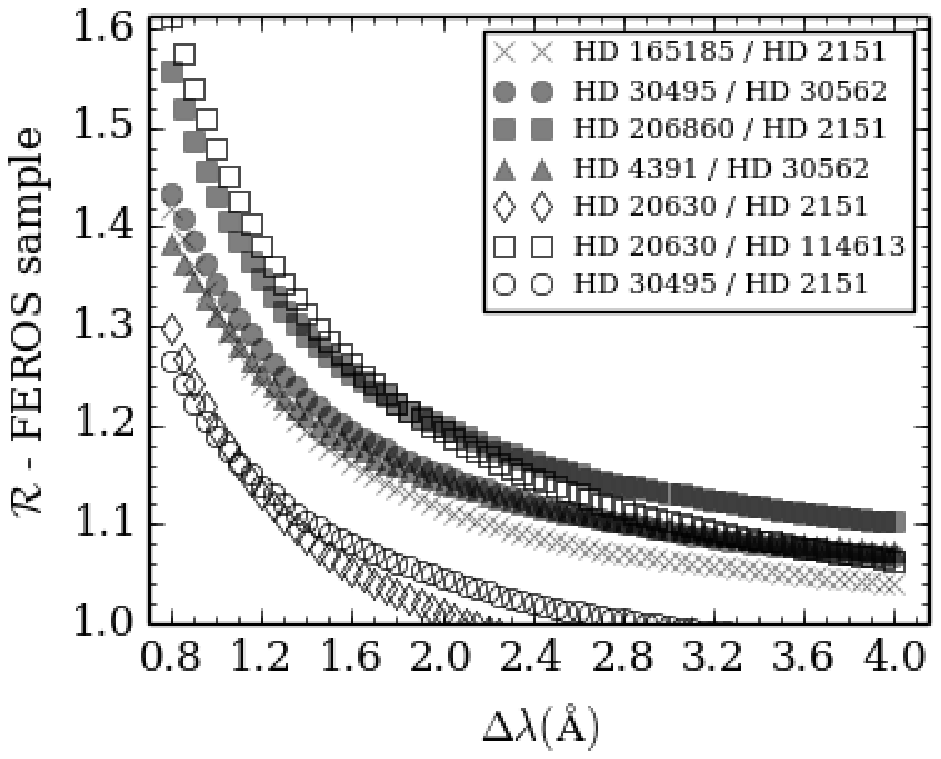}}
  \end{minipage}
   \begin{minipage}[h]{0.49\linewidth}
\resizebox{\hsize}{!}{\includegraphics{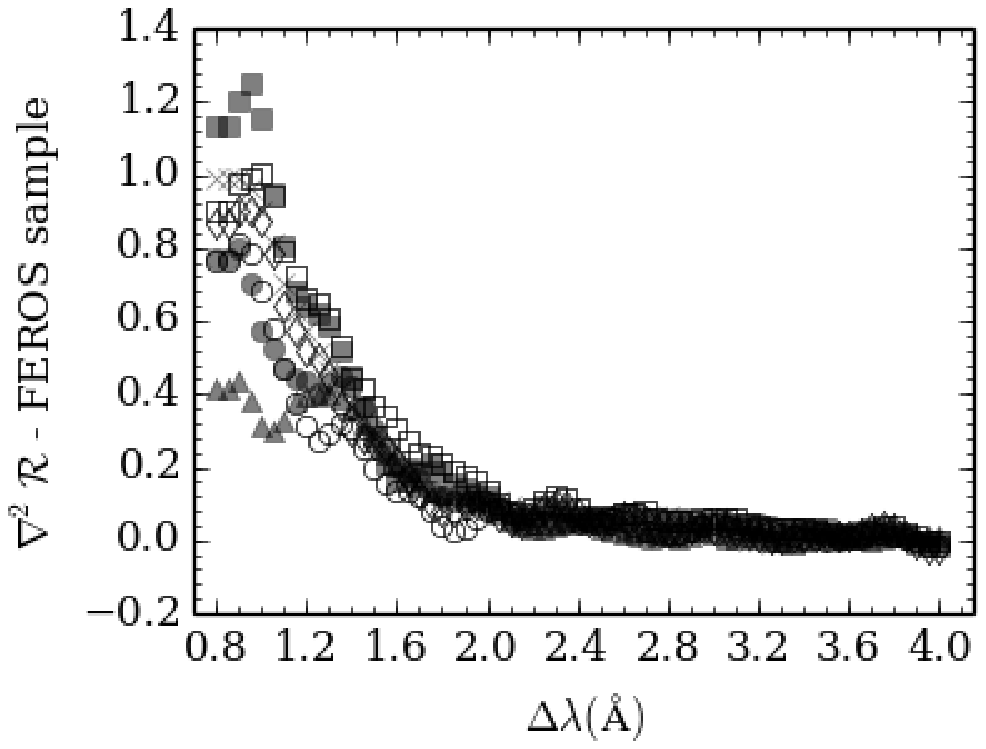}}
  \end{minipage}
\caption{Variation of the chromospheric flux component as a function of the integration bandwidth. \textit{Left:} Relation between the $\mathcal{R}$ and the integration bandwidth around T3 spectral line (FEROS database). \textit{Right:} The second order gradient of $\mathcal{R}$ for the FEROS T3 line. There is a transition which connects two different regimes around $\Delta\lambda$ = 2.0 \AA\ .}
\label{fig:razao2}
\end{figure}

The average flux is given by the five adopted reference regions ($R_i$ chosen for a particular band integration ($\Delta\lambda_{L}^j$) around the center of each Ca II IRT line. The results are shown in Fig. \ref{fig:razao2}. 
The dilution of the chromospheric component offers a basis for choosing the bandwidth for maximizing observationally the visibility of the chromospheric component. Additionally, the flux ratio differences between the pairs of stars are gradually removed as we consider the second derivative, converging to a single regime. For $\Delta\lambda$ > 2.0 \AA\, all pairs show the same dilution of the chromospheric component, in excellent agreement with a visual estimate of the visual spectra of \citet{linsky79b}. 

We applied the same procedure to the OPD subsample. Apparently instrumental effects make it more difficult to identify a smooth variation of $\mathcal{R}$ with the increased integration bandwidth. Even so, it was possible to identify the need for a wider range of $\Delta \lambda_{L}$ in the OPD spectra defined as 2.2 \AA. From these values, we calculate the total absolute flux on each Ca II IRT line based on Eq. \ref{f6}.

To convert the subsamples to the same scale, we selected stars in common and fitted a linear function relating both subsamples\footnote{For T2, the conversion was not needed since it is not present in spectra of FEROS database}:
\begin{equation}\label{eq:calibracao}
\mathrm{\langle \mathcal{F}_{L}^{FEROS} \rangle = a \langle \mathcal{F}_{L}^{OPD} \rangle + b} \equiv \mathcal{F}_{L}.
\end{equation}

\begin{figure}[!htbp]
\begin{center}
  \begin{minipage}[t]{0.8\linewidth}
    \includegraphics[width = 1\linewidth]{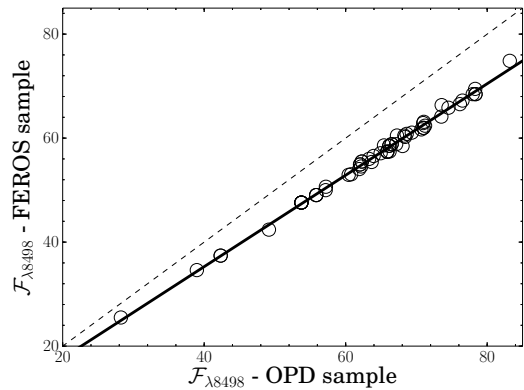}
  \end{minipage}
\end{center}
\caption[]{Adopted fits for the conversion of T1 total absolute fluxes from OPD to the FEROS database. Black dots are observations in common; dashed line is the 1:1 relation, full line is the fit.}
\label{fig:LNAFEROS}
\end{figure}

Our choice of the integration bandwidths ($\Delta\lambda_{L}^{FEROS}$ = 2.0 \AA\ and $\mathrm{\Delta\lambda_{L}^{OPD}} $ = 2.2 \AA) leads to conversion errors of $\sigma_{\mathrm{conv.}} \approx$ 0.5 $\times$ 10$^5$ erg.cm$^{-2}$s$^{-1}$, which represents typically 1\% of $\mathcal{F}_{L}$. In Fig \ref{fig:LNAFEROS}, we show the adopted fit for the final subsample conversion of the T1 line and, in Table \ref{table:conversao}, we list the values of the regression coefficients calculated in the absolute total flux regression for the Ca II IRT lines T1 and T3.
\begin{table}[!htbp]
\begin{center}
\begin{tabular}{c c c c c}
  \hline
 & $a$ & $b$ & $\sigma_{\mathrm{conv.}}$ & $R^2$ \\
\hline
$\lambda$8498 & 0.879 & 0.135 & 0.57 & 0.996 \\
$\lambda$8662 & 0.870 & -0.410 & 0.47 & 0.996 \\
  \hline
\end{tabular}
\end{center}
\caption{Coefficents of the OPD-FEROS total absolute flux conversion. The errors are in 10$^5$ erg cm$^{-2}$s$^{-1}$ units.}
\label{table:conversao}
\end{table}

Although some of our sample stars were observed over six years, it was not possible to quantify the influence of modulation of magnetic activity cycles. Therefore, any detected variability is interpreted as a result of uncertainties propagated by data reduction procedures as well as instrumental differences between each observational run. Such differences affect the accuracy of the total absolute flux scales calculated in observed spectra. Thus, we assume it to be a random effect that could be estimated by repeatability. We calculated the total absolute fluxes of the multiply observed stars and found low typical variations of $\approx$ 0.1-0.2$ \times $10$^5$  erg cm$^{-2} $s$^{-1}$ ($\sigma_{rep}$) that, coupled with the uncertainties derived from the regressions ($\sigma_{reg}$ and $\sigma_{\mathrm{parameters}}$), provide total line flux errors ($\sigma_{Total}$) of $\approx$ 1.3 to 1.4 $\times $10$^5$ erg cm$^{-2}$s$^{-1}$. The errors are given in Table \ref{table:erros}. Observed absolute line fluxes are thus seem to be very stable and we can confidently treat our sample as fully homogeneous.
\begin{table}[!htbp]
\begin{center}
\begin{tabular}{c c c c c c | c}
  \hline
& \multicolumn{5}{c}{Errors ($\times$ 10$^5$ erg cm$^{-2}$ s$^{-1}$)} \\
\hline
 line & $\sigma_{rep}$ & $\sigma_{\mathrm{param}}$ & $\sigma_{reg}$ & $\sigma_{conv.}$ & $\sigma_{Total}$ & $\sigma_{Chrom}$\\
  \hline
T1-T3  & 0.2 & 1.3 & 0.03 & 0.5 & 1.41 & 0.62\\
T1-T3  & 0.2 & 1.3 & 0.03 & 0 & 1.32 & 0.36\\
T2 & 0.1 & 1.3 & 0.03 & 0 & 1.3 & 0.32\\
   \hline
\end{tabular}
\end{center}
\caption{Derived errors in the absolute flux values. The second and third columns respectively are related to repeatability errors and propagated errors generated by observational uncertainties (\Teffe, \logge, and \fehe). The forth column shows the average of fitting errors, the fifth column shows conversion errors between the subsamples. We note that the first line denotes the errors in OPD spectra converted to FEROS database, the second line shows the propagated errors in FEROS spectra, and the third line refers to the errors in T2, available only for the OPD spectra. The sixth column shows the errors on the total absolute line fluxes $\approx$ 1.3-1.4 $\times$ 10$^5$ erg cm$^{-2}$ s$^{-1}$. The seventh column shows the chromospheric flux errors after the photospheric subtraction (see Sect. \ref{sec:act_prob}) combined with repeatability and conversion errors.}
\label{table:erros}
\end{table}

\section{Photospheric correction} \label{sec:phot_correc}

We assume that the chosen integration interval for the measurement of total flux covers completely the chromospheric contribution, but also includes some photospheric contamination that must be properly removed. The major difficulty in measuring the chromospheric flux in photospheric strong lines lies in the fact that, in most cases, the desired quantity is one or two magnitudes smaller than the total and purely photospheric flux. As emphasized in a number of papers \citep{hartmann84,rutten87,pasquini91,lyra05}, the lack of knowledge regarding the photospheric flux distribution unfortunately imposes the adoption of arbitrary methods of correction. We now briefly discuss some techniques employed to carry out this task.

The works of \citet{middelkoop82}, \citet{rutten84}, and \citet{noyes84} enable the derivation of a normalized index of chromospheric activity independent of stellar spectral type, in principle, the $\mathrm {R_{HK}^\prime}$. 

Despite the consistent results of chromospheric activity in open clusters, the strong correlation with X-rays, and rotational periods \citep{mamajek08}, such non-dimensional indices may have complicated the physical interpretation, especially in the very low-activity regime \citep{hall09}. Furthermore, as we consider stars with a wide range of \Teffe, the normalization factor may contribute adding an explicit dependence on \Teff (color), independent of intrinsic chromospheric fluxes \citep{rutten87}.

\citet{linsky79b}, inspired by \citet{wilson68}, compared stars with different levels of chromospheric activity and derived a function that represented the minimum chromospheric radiative losses given a color index (V-I). The difficulty of this method is to determine precisely which stars are adequate to establish a minimum level of magnetic activity as a function of \Teffe, or a proxy of it. Thus, it is imperative to populate the sample with a significant number of evolved stars for which we do expect systematically lower chromospheric radiative losses due to their angular momentum loss history.

It is noteworthy that this process is facilitated in samples like ours as we have the proper characterization of key parameters such as \Teff and intrinsic luminosity. On the other hand, it is clear that this method is arbitrary and strongly dependent on the sample of stars being used. It establishes a null value of chromospheric radiative loss for stars that have a minimum level of activity for a specific \Teffe. In principle, this quantity does not vanish since we expect that some residual chromospheric basal heating in evolved stars should persist \citep{schrijver87}.

\begin{figure*}[!htbp]
\begin{center}
  \begin{minipage}[t]{1\linewidth}
   \centering
  \includegraphics[width = 1 \linewidth]{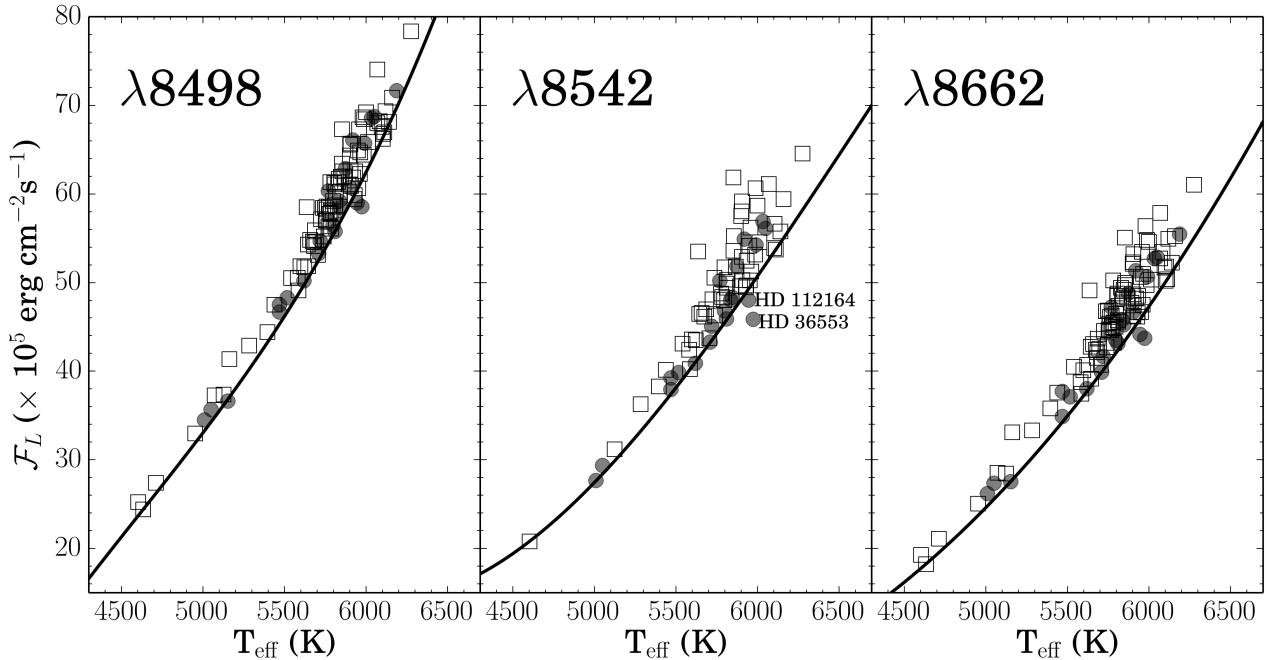}
  \end{minipage}
\end{center}
\caption{Relations between the total absolute fluxes and \Teff are shown for dwarfs (open squares) and subgiants (gray circles). The solid black lines are the lower boundary of minimum chromospheric activity represented by the more inactive subgiants. The 1.5 M$_\odot$ subgiants HD 36553 and HD 112164 are the only stars placed bellow the photospheric correction curve. The panels are divided for the three Ca II IRT lines. Left and middle panels stand for the T1 and T2 lines, and the right one ilustrates the behavior of the T3 line.}
\label{fig:term}
\end{figure*}

We expect a strong correlation between $\mathcal{F}_{L}$ and parameters closely related to the stellar structure. Thus, in Fig. \ref{fig:term}, we confirm the \Teff correlation and identify a second component that is minimally correlated to \Teffe, interpreted as the chromospheric flux. In order to remove the photospheric component, we assume a lower boundary of minimum magnetic activity given by the more inactive subgiants of our sample. \citet{lyra05} found the same behavior for H$\alpha$ line: the boundary is populated systematically by subgiants, as expected from stellar rotational evolution for this class of stars \citep{donascimento03}. From this lower boundary, which is dependent on \Teffe, we subtract the thermal component by the relation:
\begin{equation}
\mathcal{F}(\mathrm{chromospheric}) = \mathcal{F}(\mathrm{total}) - \mathcal{F}(\mathrm{photospheric}) \equiv\mathcal{F}^\prime
\end{equation}

We emphasize that our procedure is rooted in the assumption that, to first order, the influence of photospheric flux is expressed mathematically in a separable way and dependent on a single parameter, the \Teffe. In Fig. \ref{fig:term} we show the $\mathcal{F}$ vs. \Teff diagram for the Ca II triplet lines. We tested different orders of the polynomial photospheric subtraction functions from linear to cubic. As a result, due to a better agreement with the overall visual trend of $\mathcal{F}_{L}$ and simpler form, we chose a third-order polynomial. Thus, the photospheric correction was performed for each star in our sample following:
\begin{equation}\label{ajuste-fot}
\mathcal{F}^\prime = \mathcal{F}_{L}(\mathrm{total}) - (a + b\cdotp \mathrm{X} + c\cdotp \mathrm{X}^2 + d\cdotp \mathrm{X}^3),
\end{equation}
where $\mathrm{X}$ $\equiv$ \Teffe/5777. In Table \ref{table:sub}, we list the coefficients of Eq. \ref{ajuste-fot} calculated for the Ca II IRT lines.
\begin{table}[!htbp]
\begin{center}
\begin{tabular}{c c c c c}
  \hline
 Ca II Line & $a$ & $b$ & $c$ & $d$ \\
  \hline
8498 & -307.782 & 980.875 & -1063.945 & 445.561 \\  
8542 & 199.046 & -659.455 & 709.837 & -204.499 \\
8662 & 11.934 & -61.457 & 67.723 & 23.336 \\
  \hline
\end{tabular}
\end{center}
\caption{Coefficients of the third-order polynomial involved in the subtraction of the photospheric fluxes.}
\label{table:sub}
\end{table}

Two of the most massive subgiants in our sample, HD 36553 and HD 112124 (\Teff $\approx$ 5950 K, M $\approx$ 1.5 M$_\odot$, \logg $\approx$ 3.7 dex and \feh $\approx$ +0.25 dex), are very similar and consistently placed below the minimum activity boundary for all Ca II IRT lines. These stars are at the tail of our sample distribution of masses, surface gravities, and metalicities (see Fig. \ref{fig:NMARCS3}). So, considering that these subgiants are not representative of our sample and to avoid the chromospheric flux superestimation of our hottest stars, we chose to remove them from the minimum activity boundary. This case is yet another reminder of the arbitrariness of such procedure and the strong bias that can be introduced by smaller and/or inadequately populated samples.  

The photospheric subtraction procedure is well-constrained between 5100 K and 6100 K. Outside this domain, higher chromospheric flux uncertainties are expected since our fit deviates from the overall visual trend. 

\subsection{Algorithm to derive absolute chromospheric fluxes}\label{sec:steps}
We describe the necessary steps to derive the Ca II IRT chromospheric fluxes:
\begin{enumerate}
 \item Around each Ca II IRT line, calculate the observed line flux $f_{L}$ (Sect. \ref{sec:total_abs_fluxes}).
 \item For each reference region defined by Table \ref{table:regselecionadas}, obtain the observed pseudo-continumm flux $f_{C}$ (Sect. \ref{sec:total_abs_fluxes}). Then, with the help of Eqs. \ref{f7}, \ref{mcub}, and Table \ref{table:coeff}, calculate the theoretical continuum absolute flux $\mathcal{F}_C$ (which is defined as $\overline{F}\Delta\lambda_C$) for a given set of atmospheric parameters. As a result, derive the absolute total line flux $\mathcal{F}_L$ through Eq. \ref{f6}.
 \item Average all reference region estimates of $\mathcal{F}_L$ obtaining $\langle\mathcal{F}_L\rangle$. Set $\mathcal{F}_L$ $\equiv$ $\langle\mathcal{F}_L\rangle$.
  \item Calculate $\mathcal{F}^\prime_L$ after correcting the photospheric signature of $\mathcal{F}_L$ using Eq. \ref{ajuste-fot} and Table \ref{table:sub} (Sect. \ref{sec:phot_correc}).
  \item The errors of these procedures are summarized in Table \ref{table:erros}.
\end{enumerate}

\section{Uncertainty of chromospheric flux}\label{sec:act_prob}

\begin{figure}[!htbp]
\begin{center}
  \begin{minipage}[t]{1\linewidth}
   \centering
  \includegraphics[width = 1 \linewidth]{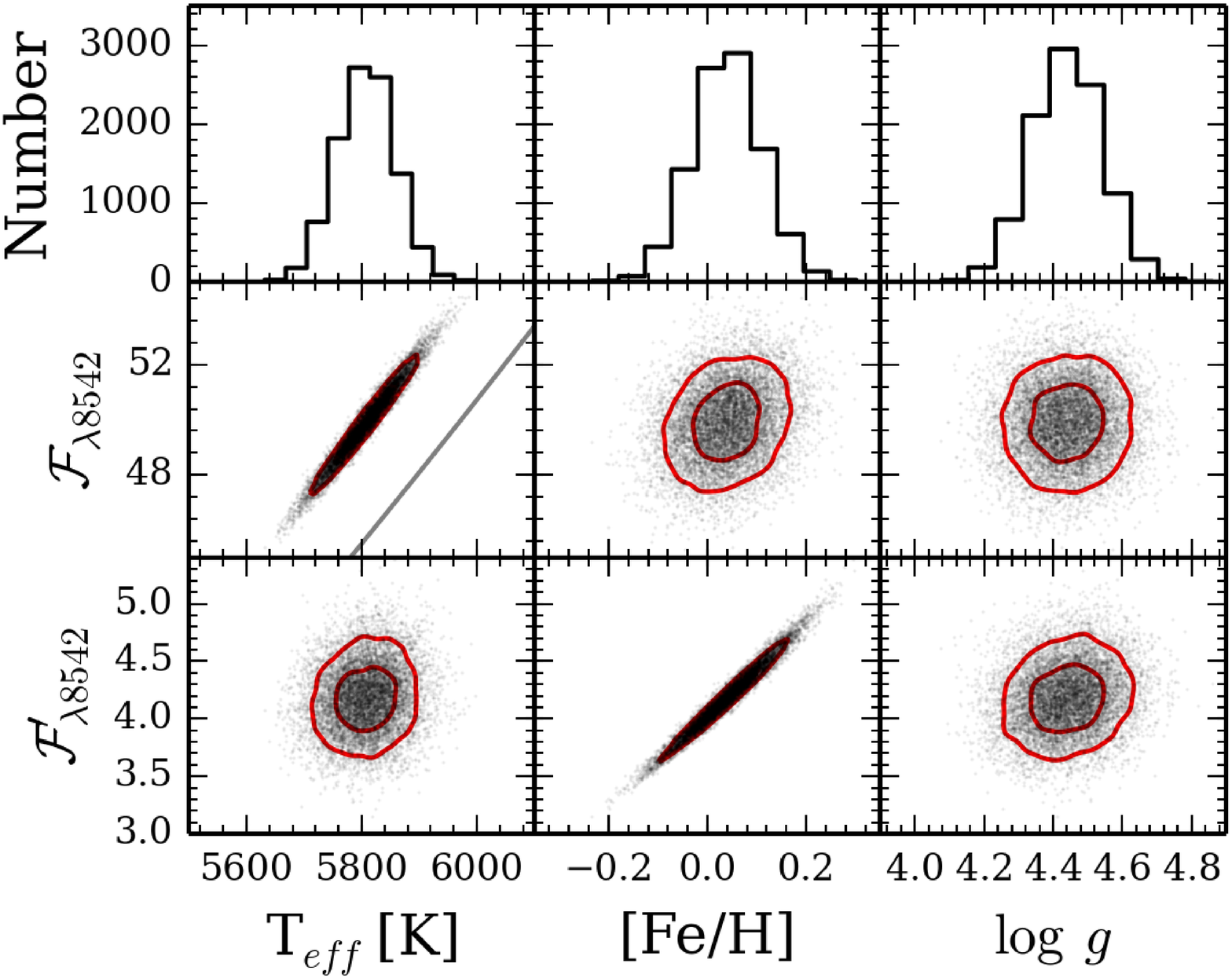}
  \end{minipage}
\end{center}
\caption{18 Sco MC output distributions of $\mathcal{F}_{\lambda8542}$ and $\mathcal{F}^\prime_{\lambda8542}$ as a function of \Teffe, \feh and \logg are shown. \textit{Upper panels:} We consider input errors of 50 K, 0.1 dex, and 0.07 dex of \Teffe, \fehe, and \logge. \textit{Middle panels:} $\mathcal{F}_{\lambda8542}$ output distribution. Solid gray line refers to the photospheric correction and the solid red lines are 1$\sigma$ and 2$\sigma$ confidence intervals. \textit{Lower panels:} $\mathcal{F}^\prime_{\lambda8542}$ correlations with atmospheric parameters.}     
\label{fig:dist_act}
\end{figure}

In order to gauge the impact of atmospheric parameters uncertainties on the chromospheric flux scale, first, we chose a representative star of our sample, the solar-twin 18 Sco \citep[\Teff = 5809 K, \feh = +0.04 dex, and \logg = 4.44 dex,][]{portodemello14}. We generated 10$^4$ MC simulations for its \Teffe, \logge, and \feh assuming Gaussian uncertainties of 50 K, 0.1 dex, and 0.07 dex, respectivelly. The generated distribution of these parameters is shown in Fig. \ref{fig:dist_act} (upper panels). Then, using the OPD spectra and the T2 line, for each MC simulation, we followed each steps described in Sect. \ref{sec:steps} and derived the output distributions of total flux $\mathcal{F}_{\lambda8542}$ (middle left panel) and photospherically corrected flux (lower panel) $\mathcal{F}^\prime_{\lambda8542}$.

In the middle and lower panels of Fig. \ref{fig:dist_act}, a sharp profile indicates a residual correlation between the related quantities. The $\mathcal{F}_{\lambda8542}$ distribution is clearly above the minimum level of activity given by the photospheric correction (solid gray line) and strongly correlated with \Teffe, which accounts for the majority of total flux variance. After subtracting the 18 Sco photospheric component, we estimated the internal chromospheric flux errors ($\sigma_{\mathcal{F}^\prime_{\lambda8542}}$) of 0.3 $\times$ 10$^5$ erg cm$^{-2}$ s$^{-1}$ which represent $\approx$ 8\% of $\mathcal{F}^\prime_{\lambda8542}$. 

The impact of increasing uncertainties on \Teff is considerably diminished by the behavior of the minimum activity boundary. For instance, 150 K errors around the 18 Sco's \Teff (which is 3$\sigma_{T_{eff}}$) increase the chromospheric flux errors by $\leq$ 7\%. Between 5400-6200 K, a simple scale shift of $\pm$\, 400 K in \Teff modifies the chromospheric flux scales slightly by an amount of $\leq$ 10\% that is, certainly, compatible with $\sigma_{\mathcal{F}^\prime_{\lambda8542}}$. This result shows the consistency of our photospheric correction inside this domain. On the other hand, \fehe, which accounts for $\approx$ 10\% of total flux variance (see Sect. \ref{sec:total_abs_fluxes}), after the photospheric subtraction turns out to be the major source of $\sigma_{\mathcal{F}^\prime_{\lambda8542}}$. In the case of higher \feh errors of 0.21 dex (3$\sigma_{\mathrm{[Fe/H]}}$), $\sigma_{\mathcal{F}^\prime_{\lambda8542}}$ increases by an amount of 300\%. Therefore, this enforces the need of an accurate determination of \feh to avoid undesirable biases in the chromospheric activity distributions. Surface gravity effects are negligible (\textless\, 3\%)\footnote{The same behaviour of the T2 line as a function of atmospheric parameters was also found for the T1 and T3 lines.}.

The role of \feh and \logg uncertainties in active stars are expected to be minimized due to the strong contrast between the photospheric and chromospheric components. Nonetheless, these effects should be determining factors for inactive stars, especially those older than the Sun, if the age-activity relation is to be extended to the end of the main-sequence.

\section{Conclusions}\label{sec:conclusions}
 
We derive absolute chromospheric fluxes in the Ca II IR triplet lines for 113 FGK
stars, from high signal-to-noise ratio (S/N) and high-resolution spectra, covering an extensive range of atmospheric parameters and chromospheric activity levels. To this purpose, we derived the first near-IR continuum calibration anchored on NMARCS LTE models of atmospheres as an explicit function of \Teffe, \fehe, and \logge. This procedure reduces the internal uncertainties in absolute continuum fluxes of Ca II IRT lines in order to enable a more consistent age-activity calibration of the solar-type stars. The internal errors on this analysis are two orders of magnitude smaller than in previous near-IR calibrations anchored in color indices. We also find excellent agreement between our absolute flux calibration and \citet{hall96} corrected from \feh bias ($\sigma$ = 0.29 $\times$ 10$^5$ erg.cm$^{-2}$s$^{-1}$). 

Through MC simulations, we find that the internal chromospheric flux error of each Ca II IRT line is 0.3 $\times$ 10$^5$ erg cm$^{-2}$s$^{-1}$ ($\approx$ 10\% of $\mathcal{F}^\prime_{L}$). Owing to the proper photospheric correction, the activity distribution width does not depend primarily on the \Teff uncertainties leaving \feh as the most important parameter in chromospheric flux variance. These new relations are especially important to bring into the same scale the chromospheric flux estimates of stars with different chemical composition, evolutionary states, and mass. Moreover, this approach provides a better understanding of the internal uncertainties and its dependencies with the atmospheric parameters. This new chromospheric flux calibration will be used in a forthcoming paper to investigate more comprehensively the age-activity relation of solar-type stars. Our reduced uncertainties enable us to explore the low-activity end of this relation, including older and less active stars. We also plan an extension of this method to late K and M dwarfs with precise ages \citep{garces11}.

Since 2003, the \textit{Radial Velocity Experiment} survey (RAVE) has been obtaining low-resolution spectra (R = 7500) in the near-infrared region (8410-8795 \AA) for thousands of stars. In addition, the GAIA mission \citep{perryman01} was successfully launched and will provide, in a unprecedented way, a 6-dimension map (positions and velocity components) for 10$^9$ stars in the Galaxy. In addition to the astrometric data, it will include a photometric and spectroscopic database of intermediate resolution (R = 7500-11500) spectra in the Ca II IRT region for millions of late-type stars. Thus an adequate calibration of chromospheric fluxes along the lines we present could potentially allow age determinations for millions of stars.

\begin{acknowledgements}
We thank the anonymous referee for helpful comments. G.F.P.M. acknowledges financial support by the CNPq grant 476909/2006-6, the FAPERJ grant APQ1/26/170.687/2004, and the CAPES post-doctoral fellowship
BEX 4261/07-0. D.L.S. acknowledges a scholarship from CNPq/PIBIC. D.L.S and L.D.F. acknowledge MSc CAPES scholarships. We thank the staff of the OPD/LNA for considerable support in the many observing runs carried out during this project. Use was made of the Simbad database, operated at the CDS, Strasbourg, France, and of NASA's Astrophysics Data System Bibliographic Services. We thank Edward Guinan, Jos\'e Dias do Nascimento Jr., and Jeffrey Hall for interesting discussions. I.R. acknowledges support from the Spanish Ministry of Economy and Competitiveness (MINECO) through grant ESP2014-57495-C2-2-R.
\end{acknowledgements}

\bibliographystyle{aa} 
\bibliography{bibli2} 

\end{document}